# An evaluation of the Australian Research Council's journal ranking

*Jerome K Vanclay*

*Southern Cross University, PO Box 157, Lismore NSW 2480, Australia*

*Tel +61 2 6620 3147, Fax +61 2 6621 2669, JVanclay@scu.edu.au*

**Abstract**

As part of its program of 'Excellence in Research for Australia' (ERA), the Australian Research Council ranked journals into four categories (A*, A, B, C) in preparation for their performance evaluation of Australian universities. The ranking is important because it likely to have a major impact on publication choices and research dissemination in Australia. The ranking is problematic because it is evident that some disciplines have been treated very differently than others. This paper reveals weaknesses in the ERA journal ranking and highlights the poor correlation between ERA rankings and other acknowledged metrics of journal standing. It highlights the need for a reasonable representation of journals ranked as A* in each scientific discipline.

## 1. Introduction

In 2008, the Australian Government announced its *Excellence in Research for Australia* (ERA) initiative to assess research quality within Australia's higher education institutions using a combination of indicators and expert review (Anon 2009). One of the indicators is a discipline-specific journal ranking (Anon 2010a), despite limitations of this approach (Butler 2003a; Bollen et al 2009; Lamp 2009; Northcott and Linacre 2010). This ranking of journals has not been universally welcomed (e.g., Peters 2008), and this contribution seeks to evaluate whether the ranking released in February 2010 is equitable across disciplines. Others have considered various aspects of expert review and esteem indicators (e.g., Donovan and Butler 2007; Genoni and Haddow 2009; Jarwal et al 2009; Atkinson 2010), so this study confines itself to an analysis of the journal ranking across disciplines. Such analyses are important to maintain the objectivity of the ERA system, because the draft ranking attracted the observation that "it is plausible to suggest that some degree of game playing may have taken place in the journal selection and allocation process. That is, some academics may have, on occasion, mixed their university specific role with their broader collegial duties" (Anderson and Tressler 2009). This study compares the ERA ranking across all disciplines and within selected disciplines, and complements other within-discipline studies (Haddow and Genoni 2010, Haslam and Koval 2010).

## 2. The distribution of journals by 2-digit FOR division

The ERA ranking allocates 20,712 journals into four quality categories, A*, A, B and C, in such a way that A* should represent the top 5% of journals, A should include the next 15%, B the next 30%, and C the remaining 50% of journals (Graham 2008). The ERA also draws on 2-digit divisions and 4-digit groups of the Australian and New Zealand Standard Research Classification (ANZSRC 2008) known as Field of Research (FOR) codes. Given that each of the FOR divisions included at least 164 journals (in the case of *05 Environmental Sciences*), and averaged 888 journals per division, it seems reasonable to assume that the distribution of journals within each 2-digit FOR division should approach the nominal 5:15:30:50. However, this is not the case for the 2010 journal list (Anon 2010b): Table 1 shows how journals are distributed across the 4 categories within 24 disciplines (23 FOR divisions plus the ERA Multidisciplinary category), using only the primary FOR code to avoid double-counting (some journals were assigned 2 or 3 FOR codes), and omitting unranked journals.

**Table 1**. Cumulative percentage of journals in ERA categories by 2-digit FOR division.

| 2-digit FOR division | A* | A | B | C |
|---|---|---|---|---|
| 01 Mathematical Sciences | 7 | 25 | 53 | 100 |
| 02 Physical Sciences | 8 | 27 | 55 | 100 |
| 03 Chemical Sciences | 8 | 24 | 49 | 100 |
| 04 Earth Sciences | 5 | 22 | 47 | 100 |
| 05 Environmental Sciences | 2 | 16 | 45 | 100 |
| 06 Biological Sciences | 6 | 18 | 41 | 100 |
| **07 Agricultural and Veterinary Sciences** | 1 | 12 | 34 | 100 |
| 08 Information and Computing Sciences | 7 | 24 | 52 | 100 |
| 09 Engineering | 7 | 24 | 53 | 100 |
| 10 Technology | 2 | 18 | 44 | 100 |
| 11 Medical and Health Sciences | 4 | 16 | 39 | 100 |
| **12 Built Environment and Design** | 15 | 39 | 60 | 100 |
| 13 Education | 3 | 16 | 48 | 100 |
| 14 Economics | 7 | 24 | 51 | 100 |
| 15 Commerce, Management, Tourism and Services | 5 | 17 | 46 | 100 |
| 16 Studies in Human Society | 4 | 21 | 51 | 100 |
| 17 Psychology and Cognitive Sciences | 4 | 20 | 50 | 100 |
| 18 Law and Legal Studies | 4 | 17 | 45 | 100 |
| 19 Studies in Creative Arts and Writing | 6 | 23 | 51 | 100 |
| 20 Language, Communication and Culture | 5 | 20 | 50 | 100 |
| 21 History and Archaeology | 3 | 23 | 59 | 100 |
| 22 Philosophy and Religious Studies | 4 | 22 | 54 | 100 |
| MD Multidisciplinary | 5 | 17 | 43 | 100 |
| **Overall** | **5** | **20** | **48** | **100** |

ERA announcements prescribed a distribution of 5:15:30:50, corresponding to cumulative percentages of 5, 20, 50, and 100%, close to what is observed across all divisions (Table 1, bottom line), but some highlighted FOR divisions depart significantly from this trend. One FOR division *12 Built Environment and Design* has significantly more A and A* journals than expected, and *07 Agricultural and Veterinary Sciences* has significantly fewer A and A* journals than expected. A $\chi^2$ test suggests that the distribution across all FOR divisions is inequitable ($\chi^2_{69}$=547, P<0.0001). The particularly high representation (15% and 24%) of A* and A journals in *12 Built Environment and Design* appears to be the successful result of an orchestrated campaign (Friedman et al 2008). Table 1 considers only the primary FOR code, overlooks secondary and tertiary FOR codes (15% of journals had 2 or 3 FOR codes in a FOR division other than the primary division), and considers only the 2-digit FOR division rather than the 4-digit FOR group. Although this simplifying assumption involves about 15% of journals, it is not sufficient to explain the inequalities in Table 1. A finer-grained analysis reveals greater inequalities.

An analysis of 4-digit FOR groups (including secondary and tertiary FOR codes) reveals that 33% of the 43 journals in *1203 Design Practice and Management* are ranked A*, whereas 0% of the 85 journals in *0705 Forestry Science* are ranked A*. These two FOR groups reflect two extremes, but are not unique, and about one quarter of the 4-digit FOR groups have no A* journals, an unhappy and inequitable situation for any field of research. Thus a more detailed examination of these two FOR groups (*0705 Forestry* and *1203 Design*) is warranted.

## 3. Why are there no A* journals in *0705 Forestry Science*?

The ERA ranking was based on expert consultation and review by the 'Learned Academies' (Australian Academy of the Humanities, Academy of Social Sciences in Australia, Australian Academy of Science, Australian Academy of Technological Sciences and Engineering; Anon 2010c), lending to the belief that it should be fair and reasonable. Perhaps the apparent discrepancy is Table 1 is warranted, because a discipline (in this case, *0705 Forestry*) does not attain the expected quality? One way to test for this possibility is to examine where well-cited forestry papers are published, and to compare the classifications of their host journals.

A Scopus search for items with the keyword 'forestry' or 'silviculture', published during the ERA census period 2003-08 returned 25198 documents. The most-cited 5% (1260) of these papers appeared in 200 journals, of which all but 29 journals hosted fewer than 10 of these highly-cited papers. These 29 journals are shown in Table 2, ranked by number of citations received. Table 2 shows that highly-cited forestry papers appear in A* journals across a wide range of FOR codes, suggesting that the ERA ranking and not the forestry discipline is underperforming. Nine of the journals carried half of the most-cited papers (624 out of the 1260 articles); three of these are classified as *0705 Forestry* journals (*Forest Ecology and Management, Canadian Journal of Forest Research, Forest Science*), all ranked A, whereas all the remaining journals were ranked A* (except *Environmental Pollution*, ranked A). Scopus, the official data provider to ERA, also provides other metrics of journal performance, and the h-index (Hirsch 2005) and SNIP (Source Normalized Impact per Paper; Moed 2010) are included in Table 2 for comparison.

**Table 2**. Journals publishing at least 10 of the most highly-cited papers in forestry from 2003-08. Horizontal line denotes the 9 journals that carry half of the most-cited articles.

| Source | FOR Code | ERA Rank | Papers | Total cites | Scopus h-index | SNIP |
|---|---|---|---|---|---|---|
| **Forest Ecology and Management** | **0705** | **A** | 248 | 9542 | 71 | 1.70 |
| *Science* | MD | A* | 58 | 6863 | 596 | 7.72 |
| *Remote Sensing of Environment* | 09 | A* | 110 | 4805 | 92 | 3.34 |
| **Canadian Journal of Forest Research** | **0705** | **A** | 77 | 2838 | 60 | 1.20 |
| *Nature* | MD | A* | 21 | 1976 | 610 | 10.69 |
| *Soil Biology and Biochemistry* | 05 | A* | 40 | 1671 | 82 | 2.10 |
| *Geophysical Research Letters* | 04 | A* | 23 | 1155 | 106 | 1.86 |
| *Environmental Pollution* | MD | A | 24 | 1012 | 75 | 2.19 |
| **Forest Science** | **0705** | **A** | 23 | 963 | 39 | 1.36 |
| **Trees - Structure and Function** | **0705** | **B** | 26 | 895 | 37 | 1.10 |
| *Atmospheric Environment* | 09 | A | 19 | 870 | 102 | 2.04 |
| *Soil Science Society of America Journal* | 07 | A | 18 | 660 | 74 | 2.07 |
| **Journal of Forestry** | **0705** | **B** | 15 | 638 | 32 | 1.20 |
| *Canadian Journal of Remote Sensing* | 09 | B | 11 | 631 | 29 | 0.61 |
| *Conservation Biology* | 05 | A* | 15 | 629 | 99 | 2.69 |
| *Biomass and Bioenergy* | 09 | A | 12 | 625 | 51 | 2.29 |
| *Plant Journal* | 06 | A* | 12 | 594 | 124 | 2.77 |
| *Proc. National Academy of Sciences* | MD | A* | 10 | 570 | 390 | 3.61 |
| *Climatic Change* | 04 | A | 14 | 560 | 66 | 1.86 |
| *Journal of Hydrology* | 04 | A* | 12 | 556 | 76 | 2.19 |
| **Scandinavian Journal of Forest Research** | **0705** | **B** | 11 | 522 | 31 | 0.86 |
| *Plant Physiology* | 06 | A* | 13 | 498 | 132 | 2.92 |
| *International Journal of Remote Sensing* | 09 | B | 15 | 493 | 71 | 1.17 |
| *Landscape and Urban Planning* | 09 | A | 13 | 484 | 43 | 2.17 |
| *Applied and Environmental Microbiology* | 06 | A* | 11 | 464 | 159 | 2.15 |
| *Environmental Science and Technology* | MD | A* | 13 | 443 | 152 | 2.62 |
| *Journal of Experimental Botany* | 06 | A | 10 | 404 | 88 | 2.16 |
| *IEEE Trans. Geoscience &Remote Sensing* | 04 | A | 10 | 391 | 86 | 3.19 |
| *Ecological Modelling* | 05 | A | 10 | 374 | 69 | 1.44 |

Of the 894 highly-cited papers listed in Table 2, about half (47%) appear in journals classified as *0705 Forestry* (Table 3). Amongst the papers appearing in forestry (0705) journals, 88% appear in four journals ranked as A, whereas amongst papers published outside of forestry (with FOR codes other than 0705), 71% appear in 12 journals all ranked as A* (Table 3). Fisher's exact test for the 2x2 contingency table (0705 vs other; A* vs other) supports the notion that forestry journals are undervalued (P=0.023).

**Table 3**. ERA-ranking of journals publishing the most-cited forestry papers 2003-08

| ERA category | No of journals | | | | No of papers | | | | No of citations received | | | |
|---|---|---|---|---|---|---|---|---|---|---|---|---|
| | A* | A | B | Tot | A* | A | B | Tot | A* | A | B | Tot |
| MD | 4 | 1 | | 5 | 102 | 24 | | 126 | 9852 | 1012 | | 10864 |
| 04,05,06 | 7 | 4 | | 11 | 126 | 44 | | 170 | 5567 | 1729 | | 7296 |
| 07 | | 4 | 3 | 7 | | 366 | 52 | 418 | | 14003 | 2055 | 16058 |
| 09 | 1 | 3 | 2 | 6 | 110 | 44 | 26 | 180 | 4805 | 1979 | 1124 | 7908 |
| Subtotal | 12 | 12 | 5 | 29 | 338 | 478 | 78 | 894 | 20224 | 18723 | 3179 | 42126 |

There is a tendency (as expected) for the ERA rank to decline with position in Table 2, with more A* journals at the top, and fewer at the bottom of the table. This trend is more evident when ERA rank (expressed as the percentile at the class midpoint, A*=0.975, A=0.875, B=0.65) is plotted against journal rank based on total citations (Figure 1, *Forest Ecology and Management* =1, *Ecological Modelling* =29). When plotted in this way, it is clear that some journals (those categorised by ERA as multidisciplinary, *04 Earth Sciences*, *05 Environmental Sciences* and *06 Biological Sciences*) are more likely to be ranked A* than those categorised as *07 Agriculture and Veterinary Sciences* or *09 Engineering*, and that they 'hold their value better' with increasing rank (Figure 1).

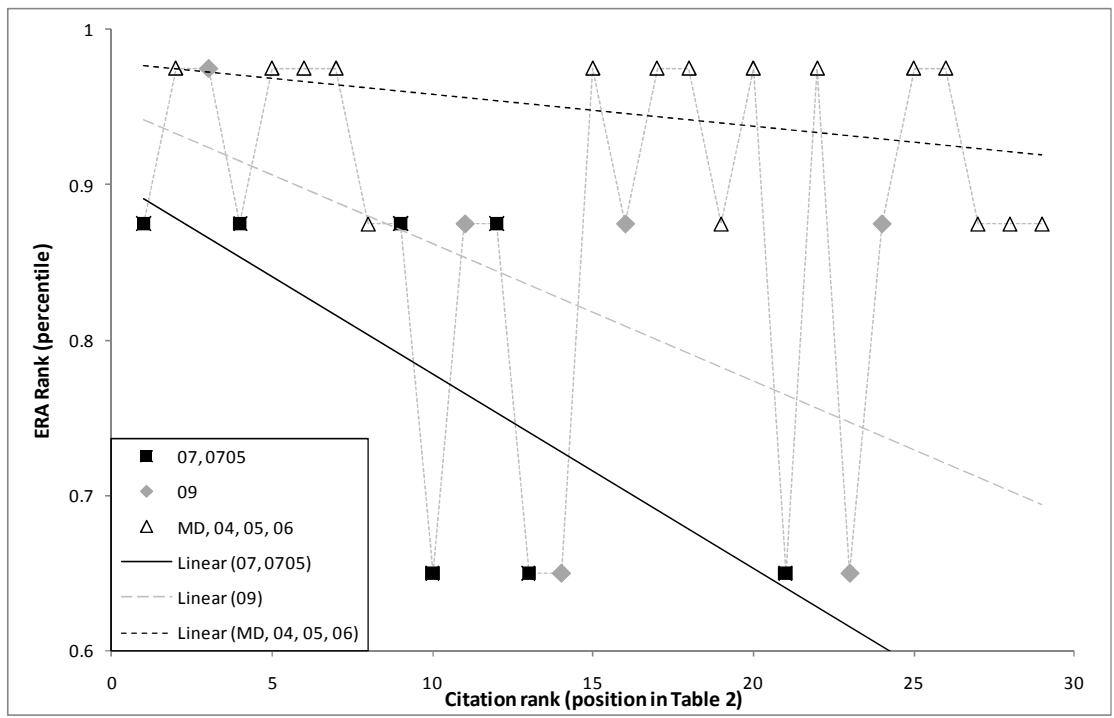

**Figure 1**. ERA rank (percentiles, A*=0.975, A=0.875, B=0.65) versus citation rank (total citations to journals bearing papers with keywords 'forestry' or 'silviculture', see Table 2). Journals categorised as 0705 Forestry (black squares, solid line) have a lower ERA rank despite a comparable citation count.

## 4. Why are there so many A* journals in *1203 Design Practice and Management*?

The discipline *1203 Design* is more difficult to analyse, because it lacks unique and distinctive keywords such as "silviculture". However, a Scopus search of the 42 journals classified as *1203 Design* reveals that three distinctive keywords were prevalent (ergonomics, biomechanics, and kinematics), and that the bulk of publications were classified by Scopus into four subject categories (Engineering, Computer Science, Social Science, Arts and Humanities). A subsequent search constrained to these subject categories and keywords, and the ERA census period 2003-08 recovered more papers classified as engineering (09) than *1203 Design* (Table 4), but confirmed similar journal rankings within these two disciplines, and illustrated how this search approach (5% of the most highly cited papers) successfully identifies many A*-ranked journals.

**Table 4**. Journals publishing at least 10 of the most highly-cited papers with keywords ergonomics, biomechanics, or kinematics. Horizontal line denotes the 12 journals that carry half of the most-cited articles.

| Source | Papers | Total cites | FOR code | ERA rank | Scopus h-index | SNIP |
|---|---|---|---|---|---|---|
| *Biomaterials* | 77 | 3935 | 09 | A* | 133 | 4.06 |
| *Journal of Biomechanical Engineering* | 75 | 2824 | 09 | A | 58 | 1.19 |
| *Science* | 60 | 6465 | MD | A* | 596 | 7.72 |
| *Proceedings of the National Academy of Sciences* | 54 | 3494 | MD | A* | 390 | 3.61 |
| *Annals of Biomedical Engineering* | 46 | 1794 | 09 | A* | 53 | 1.14 |
| *Journal of Biomedical Materials Research - Part A* | 44 | 1709 | 09 | A* | 49 | 1.30 |
| *Nature* | 44 | 4033 | MD | A* | 610 | 10.69 |
| **Ergonomics** | **33** | **1020** | **1203** | **A** | **45** | **1.49** |
| **Journal of Mechanical Design** | **32** | **1247** | **1203** | **A*** | **46** | **3.16** |
| *International Journal of Plasticity* | 20 | 822 | 09 | A | 53 | 4.10 |
| *IEEE Transactions on Robotics and Automation* | 19 | 1129 | 09 | B | | |
| *Mechanism and Machine Theory* | 19 | 680 | 09 | A | 36 | 3.45 |
| **Applied Ergonomics** | **18** | **484** | **1203** | **A*** | **32** | **1.71** |
| *IEEE Transactions on Robotics* | 17 | 572 | 09 | A* | 30 | 3.89 |
| *Acta Biomaterialia* | 16 | 684 | 09 | A | 26 | 1.57 |
| *IEEE Transactions on Biomedical Engineering* | 16 | 661 | 09 | A* | 76 | 2.00 |
| *Human Movement Science* | 14 | 437 | 09 | B | 35 | 1.75 |
| *J. Biomed. Materials Research - B Appl. Biomaterials* | 12 | 431 | 09 | A | 31 | 1.04 |
| *Annual Review of Biomedical Engineering* | 11 | 715 | 09 | A | 55 | 1.40 |
| *Biomechanics and Modeling in Mechanobiology* | 11 | 397 | 09 | C | 16 | 1.22 |
| *Journal of Fluid Mechanics* | 11 | 365 | 09 | A* | 90 | 2.33 |
| *Journal of the Mechanics and Physics of Solids* | 11 | 505 | 09 | A* | 79 | 2.57 |
| *ACM Transactions on Graphics* | 10 | 606 | 08 | A* | 76 | 7.67 |

Because the keyword search reported in Table 4 returned relatively few journals classified as *1203 Design*, it is worth examining citation rates within all journals classified as *1203 Design* (Lamp 2010a). Table 5 shows the journals with at least 10 of the papers that were amongst the 312 most-cited 5% of articles during 2003-08 in the FOR *1203 Design*. Many of these journals are ranked A*, and half of these papers were published in two journals, *Journal of Mechanical Design* , ranked A* and *Computer Aided Design* ranked A (A similar analysis for *0705 Forestry* also reveals that half of the most-cited articles appear in two journals, *Forest Ecology and Management* and *Agricultural and Forest Meteorology*, both ranked A). Clearly, some of the *1203 Design* journals warrant an A* classification as they carry contributions comparable to A*-ranked peers in the 09 Engineering discipline. However, not all the A*-ranked journals in *1203 Design* rate so well: for instance, Scopus records that over 66% of papers published during 2003-08 in the A* journals *Leonardo* and *Winterthur Portfolio* remain uncited (in sources visible to Scopus), suggesting that these journals are in a different league to the journals listed in Table 4, all of which have non-citation rates below 30%. Others (Starbuck 2005, Oswald 2007, Singh et al. 2007) have noted a high frequency of uncited papers in other prestigious journals.

**Table 5**. Selected journals classified as *1203 Design Practice and Management* and publications during 2003-08.

| Journals with ≥10 of the most-cited 5% of articles | ERA rank | Amongst top 5% | | Scopus | |
|---|---|---|---|---|---|
| | | Papers | Cites | h-index | SNIP |
| *Journal of Mechanical Design* | A* | 82 | 3389 | 46 | 3.16 |
| *Computer Aided Design* | A | 81 | 3189 | 51 | 3.57 |
| *Journal of Product Innovation Management* | A* | 39 | 1484 | 47 | 2.91 |
| *Ergonomics* | A | 33 | 1055 | 45 | 1.49 |
| *Interacting with Computers* | B | 20 | 691 | 27 | 2.38 |
| *Environment and Planning B: Planning and Design* | A* | 12 | 479 | 33 | 1.25 |
| *Applied Ergonomics* | A* | 17 | 474 | 32 | 1.71 |
| *Design Studies* | A* | 11 | 454 | 30 | 2.37 |
| **A* journals with ≤11 of the most-cited 5% articles** | | In total | | Scopus | |
| | | Papers | Cites | h-index | SNIP |
| *Design Studies* | A* | 189 | 1490 | 30 | 2.37 |
| *Research in Engineering Design* | A* | 87 | 763 | 27 | 2.28 |
| *Journal of the Textile Institute* | A* | 242 | 452 | 16 | 0.75 |
| *Leonardo* | A* | 393 | 243 | 6 | 0.19 |
| *Journal of Design History* | A* | 126 | 130 | 5 | 1.12 |
| *Design Issues* | A* | 122 | 114 | 4 | 0.72 |
| *Fashion Theory - Journal of Dress Body and Culture* | A* | 169 | 114 | 4 | 0.42 |
| *Digital Creativity* | A* | 96 | 75 | 3 | 0.37 |
| *J. Textile Institute Part 1: Fibre Science and Textile Tech.* | A* | 48 | 54 | | |
| *Winterthur Portfolio* | A* | 44 | 10 | 3 | 1.13 |

In both cases (*0705 Forestry* and *1203 Design*), the two top journals carry more than half of the papers that are frequently cited (in the top 5% of the most-cited papers). In both cases, frequently-cited papers tend to average about 37 citations/paper (as at October 2010, to papers published 2003-08). But in the case of *1203 Design*, half of these citations accrued to journals ranked by ERA as A*, whereas in *0705 Forestry* there are no A* journals, so 82% of citations accrued to journals ranked as A (Table 6). This discrepancy warrants further examination.

**Table 6**. Frequently-cited papers and the ERA rankings of the journals in which they appeared.

| ERA rank | 0705 Forestry Science | | | | 1203 Design Practice and Management | | | |
|---|---|---|---|---|---|---|---|---|
| | Papers | Cites | % | Cites /paper | Papers | Cites | % | Cites /paper |
| A* | | | | | 173 | 6,650 | 54% | 38 |
| A | 819 | 30,144 | 82% | 37 | 127 | 4,830 | 40% | 38 |
| B | 133 | 4,627 | 13% | 35 | 20 | 691 | 6% | 35 |
| C | 52 | 1,703 | 5% | 33 | 1 | 22 | 0% | 22 |
| Total | 1004 | 36,474 | | 36 | 321 | 12,193 | | 38 |

## 5. Comparing two Fields of Research: *0705 Forestry* and *1203 Design*

Tables 1 and 6 suggest some weaknesses in the ERA ranking, so further examination with independent yardsticks is warranted. One possibility is the Hirsch (2005) h-index, devised for individuals, but which can also be applied to journals (Braun et al 2006, Vanclay 2008a, Harzing and van der Wal 2009). Norris and Oppenheim (2010) have shown that individual ranking by peer assessment is generally well-correlated with the h-index. The journal h-index used here was derived from Scopus data 1996-2009 (via SCImago 2010): the long time interval 1996-2009 avoids some issues of size dependence (van Raan 2006), but disadvantages newly-established journals. Nonetheless, the graph of h-index versus the ERA category is insightful (Figure 2), and similar to the corresponding graph of ERA rank versus SCImago Journal Rank (Gonzalez-Pereira et al 2010).

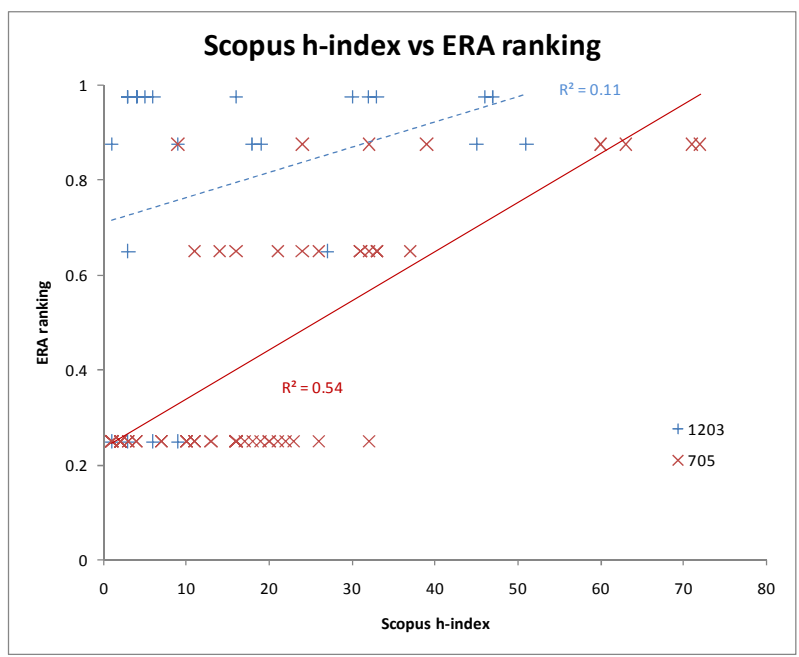

**Figure 2**. Scopus h-index versus ERA category (A*=0.975, A=0.875, B=0.65, C=0.25)

Figure 2 offers several insights. The classification of *1203 Design* journals (+, dashed line) shows a low correlation (0.11) with h-index, and examples of both low and high h-indices may occur in any of the 4 categories. In contrast, the classification of the *0705 Forestry* journals (×, solid line) shows a higher correlation (0.54) with h-index. Of greater concern is that several A-ranked journals within *0705 Forestry* have higher h-indices that A* journals within *1203 Design*, and the linear trend indicates that *0705 Forestry* journals tend to have a substantially higher h-index across all categories (i.e., the solid line is further to the right than the dashed line). The two trends differ significantly ($F_{2,75}=26.2$, $P<0.0001$). Figure 2 is based on data drawn from Scopus, but a similar relationship can be derived for journals not listed in Scopus by using data from Google Scholar (albeit with data of more variable quality, Bar-Ilan 2008).

It is possible that the h-index offers a more favourable view of some disciplines, and a less favourable view of others, so it is appropriate to consider an alternative yardstick. Scopus (the database provider to ERA) offers their own measure of journal quality, the SNIP (Source Normalized Impact per Paper; Moed 2010), which offers some independence as it was published after public submissions on the ERA ranking closed. The SNIP measures a journal's contextual citation impact, taking into account characteristics of its field, the frequency with which authors cite other papers in their reference lists, the rate of maturation of citation impact, and the extent to which a database used for the assessment covers the field's literature (Moed 2010). The correlation between ERA category and SNIP (Figure 3) is consistent with the pattern observed with the h-index.

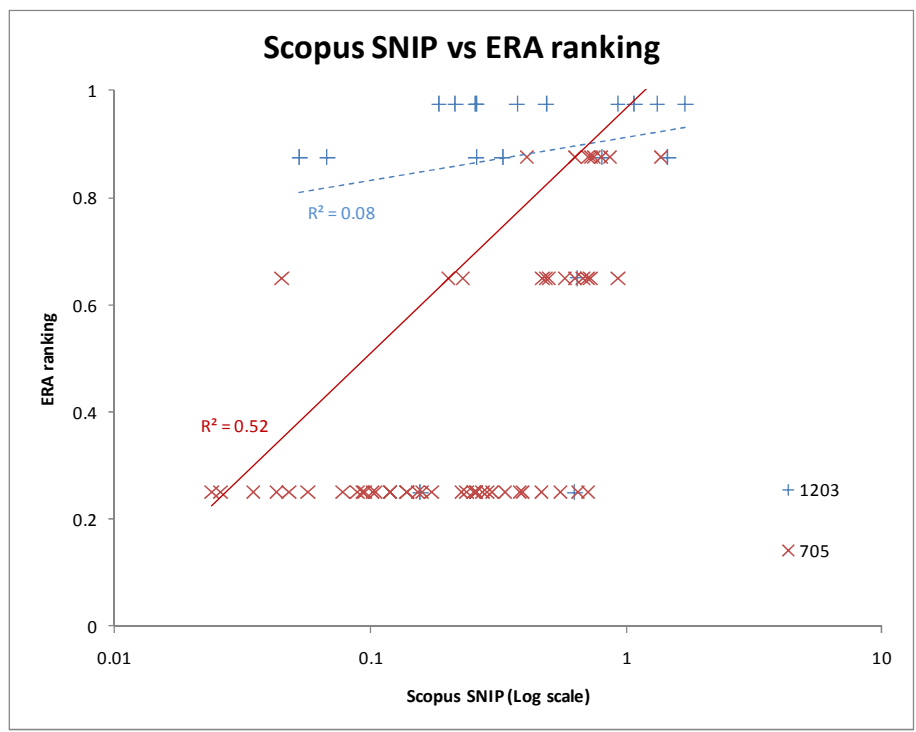

**Figure 3**. Scopus SNIP (2009) versus ERA category

Figure 3 reaffirms the weak correlation (0.08) between the ERA ranking and other indicators of journal quality within *1203 Design* journals, and illustrates that journals with low and high SNIPs appear in all the ERA categories devised for the FOR group *1203 Design*. Figure 3 also reproduces the higher correlation (0.52) between the ERA ranking and the SNIP for journals within the FOR group *0705 Forestry*, and again illustrates several examples of A-ranked journals within *0705 Forestry* that have higher SNIPs than A* journals in *1203 Design*. The two trends differ significantly ($F_{2,75}$=24.8, P<0.0001).

One advantage of the h-index is that it can be computed for all journals, and is not confined to those for which Scopus has computed a SNIP. The similarity between Figures 2 and 3 suggest that h-index offers a reasonable basis for comparisons across journals and FOR groups. There is an increasing body of evidence indicating that the h-index is a good measure of journal impact in both science and commerce (Harzing and van der Wal 2008, 2009; Imperial and Rodríguez-Navarro 2007; Vanclay 2008b). However, users should be aware that Google Scholar h-indices tend to be higher, and more subject to spurious data, than h-indices derived from Scopus and Web of Science (Bar-Ilan 2008, Meho and Rogers 2008).

## 6. Comparing Scopus SNIP and journal rank across 2-digit FOR divisions

These trends observed for FOR groups *0705 Forestry* and *1203 Design* are not unique: a graph of Scopus SNIP versus ERA category for all disciplines reveals a similar lack of discrimination across categories (Figure 4). The stepped lines in Figure illustrate the bounds that would result if ranking was based solely on the SNIP within each 2-digit FOR division (i.e., the lines represent the extremes of the 50$^{th}$, 80$^{th}$, and 90$^{th}$ percentiles from each FOR division), and thus illustrates that many journals are classified in a way inconsistent with SNIP scores.

The large number of symbols in Figure 4 makes interpretation difficult, so it is useful to examine the mean SNIP score for each ERA category (Table 7) to further evaluate these trends. Because of the non-normal distribution of these values, Table 7 reports the log-average SNIP (exponent of the average of log(SNIP)).

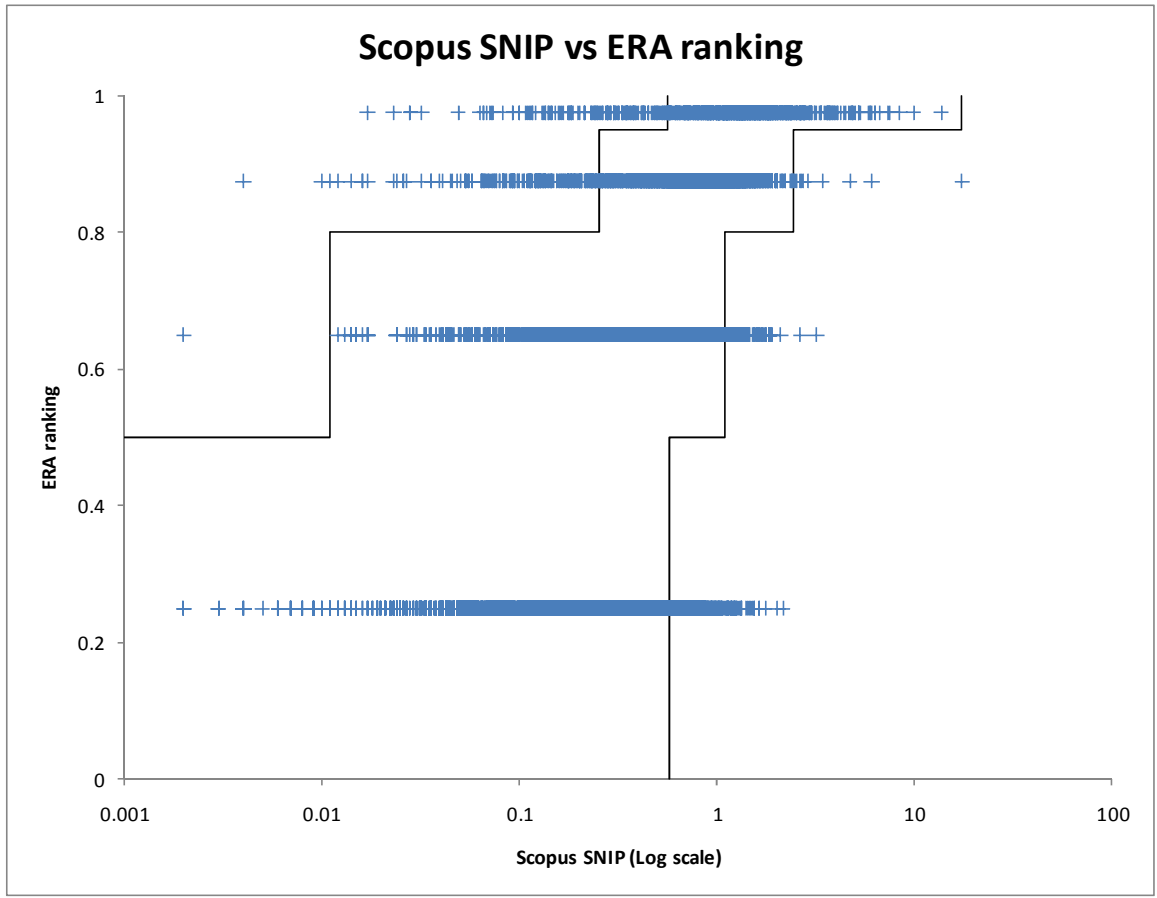

**Figure 4.** Scopus SNIP (2009) versus ERA category for all Scopus-listed journals (n=9118).

Table 7. Log-average SNIP score for each ERA category

| Discipline (2-digit FOR division) | A* | A | B | C | Average |
|---|---|---|---|---|---|
| 01 Mathematical Sciences | 2.0 | 1.2 | 0.9 | 0.5 | 1.0 |
| 02 Physical Sciences | 3.7 | 1.6 | 0.8 | 0.4 | 0.8 |
| 03 Chemical Sciences | 2.2 | 1.5 | 0.9 | 0.4 | 0.7 |
| 04 Earth Sciences | 2.3 | 1.6 | 1.0 | 0.5 | 0.8 |
| 05 Environmental Sciences | 2.5 | 1.4 | 1.1 | 0.5 | 0.8 |
| 06 Biological Sciences | 3.1 | 1.6 | 1.0 | 0.5 | 0.8 |
| **07 Agricultural and Veterinary Sciences** | **2.6** | **1.3** | **0.8** | **0.4** | **0.7** |
| 08 Information and Computing Sciences | 3.9 | 2.2 | 1.2 | 0.8 | 1.4 |
| 09 Engineering | 2.4 | 1.7 | 0.9 | 0.4 | 0.9 |
| 10 Technology | 2.0 | 1.7 | 0.9 | 0.5 | 0.9 |
| 11 Medical and Health Sciences | 2.6 | 1.4 | 0.9 | 0.4 | 0.7 |
| **12 Built environment and design** | **0.6** | **0.8** | **0.5** | **0.3** | **0.6** |
| 13 Education | 1.9 | 1.2 | 0.8 | 0.5 | 0.8 |
| 14 Economics | 2.7 | 1.3 | 0.7 | 0.4 | 0.9 |
| 15 Commerce, Management, Tourism and Services | 2.9 | 1.4 | 0.7 | 0.4 | 0.8 |
| 16 Studies In Human Society | 1.7 | 1.0 | 0.6 | 0.4 | 0.7 |
| 17 Psychology and Cognitive Sciences | 3.0 | 1.6 | 1.0 | 0.5 | 0.9 |
| 18 Law and Legal Studies | 0.5 | 0.3 | 0.2 | 0.2 | 0.3 |
| 19 Studies in Creative Arts and Writing | 0.6 | 0.3 | 0.2 | 0.3 | 0.4 |
| 20 Language, Communication and Culture | 0.9 | 0.5 | 0.4 | 0.3 | 0.5 |
| 21 History and Archaeology | 0.5 | 0.4 | 0.4 | 0.2 | 0.4 |
| 22 Philosophy and Religious Studies | 0.7 | 0.4 | 0.3 | 0.3 | 0.4 |
| MD Multidisciplinary | 2.1 | 1.0 | 0.6 | 0.3 | 0.5 |
| Average | 2.0 | 1.2 | 0.8 | 0.4 | 0.7 |

Overall, the average SNIP scores shown on the bottom line of Table 7 are consistent with the intention of the ERA, but this pattern is not evident within all the 2-digit FOR divisions. Ten cells in Table 7 have been shaded to illustrate departures from the expected trend. The five cells shaded light grey/yellow have higher than expected SNIP means, and the dark grey/green shading indicates five cells with lower than expected SNIP means. The high scores for the A* category in *02 Physical Sciences*, *06 Biological Sciences* and *07 Agricultural and Veterinary Sciences* suggests that the selection of journals was exceptionally rigorous and that inclusion of additional journals in the A* category may well be warranted. In the case of *07 Agricultural and Veterinary Sciences*, both Tables 1 and 2 suggest systematic under-representation of journals in the A* category. A* journals in the 2-digit FOR division *12 Built Environment and Design* exhibit a low average SNIP (dark grey/green shading), and this, coupled with the over-representation suggested in Table 1, suggests that the selection of journals for this category has been less rigorous. Similarly, the dark shading (green, lower than expected SNIP) for the A* category and light shading (yellow, higher than expected SNIP) for the C category suggests a less rigorous approach to the classification of journals in the 2-digit FOR divisions *19 Studies in Creative Arts and Writing*, and *21 History and Archaeology.*

## 7. Does it matter?

As a group, academics tend to be rational and respond to incentives and performance measures (Butler 2003b, 2005). During the 1990s, Australian government incentives rewarded quantity not quality, and stimulated increased publication by Australian academics in lower-impact journals, without a corresponding increase in the high impact journals (Butler 2003b). The government announced a new Research Quality Framework (RQF) in 2005, and the journal rankings (Butler 2008) that took shape during 2006 influenced publication patterns, refocusing the flow of publications into high-impact journals (Figure 5). In 2007, a new government abandoned the RQF in favour of ERA, and the journal rankings released in early 2010 appear to have stimulated a renewed emphasis on publication in A*-ranked journals. Figure 4 shows the proportion of Scopus-listed publications affiliated with one or more Australian universities and that were published in journals ranked A* by the ERA in February 2010. It is premature to attribute the kinks in this trend to the RQF in 2006 and the ERA in 2010, but the trend is suggestive of behavioural change by university academics.

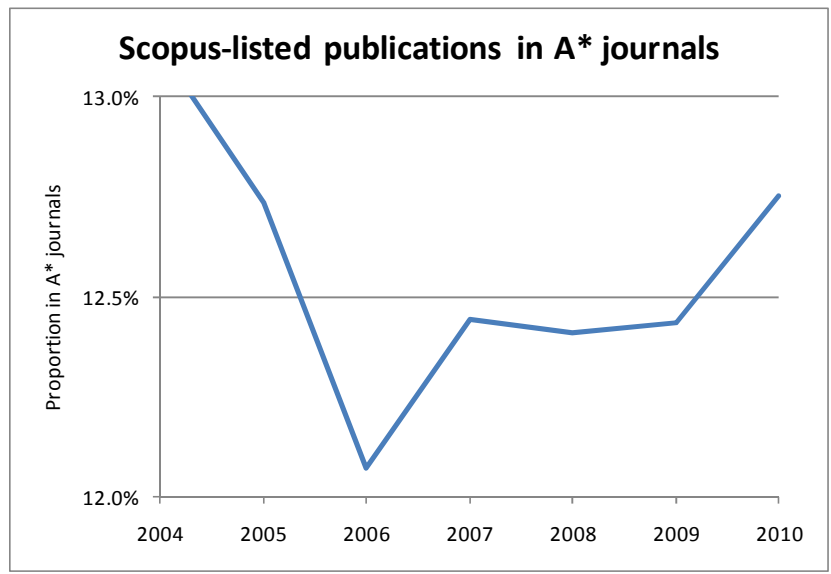

**Figure 5**. Scopus-listed journal articles published by Australian academics.

Australian academics are under considerable pressure to publish in A*-ranked journals, and to achieve the threshold of 50 publications in selected FOR groups. When the ERA creates FOR groups with no A*-ranked journals, this creates a conflict for academics. This conflict may have real and serious practical consequences. For instance, Scopus data reveals that Australian academics contribute about 5% of the papers categorized as *0705 Forestry* worldwide. In selected fields, for instance research concerning the genus *Eucalyptus*, Australian academics have an even higher impact, contributing 40% of all B- and C-ranked publications, and 60% of A-ranked publications worldwide (in journals categorised *0705 Forestry*). Closer to home, Australian academics contribute about half the articles published in the national C-ranked journal *Australian Forestry* that received by all members of the professional Institute of Foresters of Australia. The professionalism of forestry in Australia, and the viability of this journal, may be threatened if Australian academics are motivated to divert their contributions elsewhere. Science will suffer if the effort to improve research excellence in Australia motivates Australian researchers to publish their work in generic A*-ranked multidisciplinary journals instead of in disciplinary journals that constitute the mainstream of their science.

There is some evidence that scientific contributions are best reviewed within their own discipline, where reviews may be the most stringent. Issues such as the Schön affair (Beasley et al 2002) beg the question whether prominent journals such as *Nature* and *Science* are more prone to inadequate review. The incidence of errata and retractions in these journals is higher than in disciplinary journals (Table 8): this is not necessarily indicative of inadequate reviewing, and may also reflect a stringent approach to errors and retractions. However, Table 8 does lend support the notion that within-discipline publication is rigorous, and thus that the ERA should provide A* journals within each 4-digit FOR group.

**Table 8**. Errata and retractions published in Scopus-listed journals 2003-08.

| Class of journal (Scopus) | Total papers | Errata & retractions Papers | Percent |
|---|---|---|---|
| *Nature* and *Science* | 29,659 | 739 | 2.5% |
| Multidisciplinary | 98,766 | 1307 | 1.3% |
| Environmental & Agriculture | 1,107,472 | 6029 | 0.5% |
| Forestry (ERA FOR group 0705) | 20,086 | 137 | 0.7% |

## 8. Conclusion

This paper has attempted to test a series of hypotheses regarding the ERA initiative. It has examined the assumption that the ERA journal classification is fair and equitable across all disciplines (rejected, $\chi^2_{69}$=547, P<0.0001), that the FOR group 0705 Forestry has been treated fairly and equitably (rejected, Fisher's exact test, P=0.023), that all the A*-ranked journals in 1203 are of equally high standing (rejected, over half of the A* journals have none of the most-cited papers), and that the ERA classification for both *0705 Forestry* and *1203 Design* exhibit comparable trends with other measures such as h-index and SNIP (rejected, trends differ, $F_{2,75}$=24.8, P<0.0001).

It appears that the present ERA classification lacks sufficient rigour in terms of the relative numbers of journals in each category, and in terms of other independent indicators of quality (such as h-index and SNIP). These discrepancies detract from the credibility and impartiality of the ERA classification, and further revision appears warranted. These limitations of the ERA are likely to have a detrimental effect in disciplines that lack sufficient journals ranked as A*.

ERA should re-examine the distribution of journals within and between each FOR group; should consider the merits of replacing the four quality categories with a continuum defined by a metric such as SNIP or h-index; and should consider abandoning a journal-based approach in favour of an article-based approach (e.g., citations accruing to each paper, possibly weighted cf. PageRank, Brin and Page 1998). Either alternative would be preferable to the current categorical approach, because it would align authors, publishers and institutions in fostering public access to, and uptake of research. Such revision is important and urgent, because the current ERA is likely to be detrimental to several scientific disciplines.